\documentclass{article}
\usepackage{spconf-new,amsmath,graphicx}
\usepackage{svg}
\usepackage{booktabs}
\usepackage{graphviz}
\usepackage{algorithm}
\usepackage{algpseudocode}
\usepackage{soul}
\usepackage{caption}
\usepackage{hyperref}

\title{GPU-Accelerated WFST Beam Search Decoder for CTC-based Speech Recognition}

\name{Daniel Galvez, Tim Kaldewey}
\address{NVIDIA \\ 2788 San Tomas Expressway Santa Clara, CA. 95050}
\copyrightnotice{979-8-3503-0689-7/23/\$31.00~\copyright2023 IEEE}

\begin{document}
%
\maketitle
\begin{abstract}
While Connectionist Temporal Classification (CTC) models deliver state-of-the-art accuracy in automated speech recognition (ASR) pipelines, their performance has been limited by CPU-based beam search decoding. We introduce a GPU-accelerated Weighted Finite State Transducer (WFST) beam search decoder compatible with current CTC models. It increases pipeline throughput and decreases latency, supports streaming inference, and also supports advanced features like utterance-specific word boosting via on-the-fly composition. We provide pre-built DLPack-based python bindings for ease of use with Python-based machine learning frameworks at \href{https://github.com/nvidia-riva/riva-asrlib-decoder}{https://github.com/nvidia-riva/riva-asrlib-decoder}. We evaluated our decoder for offline and online scenarios, demonstrating that it is the fastest beam search decoder for CTC models. In the offline scenario it achieves up to 7 times more throughput than the current state-of-the-art CPU decoder and in the online streaming scenario, it achieves nearly 8 times lower latency, with same or better word error rate.


\end{abstract}
\begin{keywords}
Automatic speech recognition, decoder,
WFST, parallel computing
\end{keywords}
\section{Introduction}
\label{sec:intro}

Beam search is an algorithm used during inference in ASR as an alternative to greedy search. Unlike greedy search, which selects the most likely output token determined by the acoustic model at every time step, beam search maintains a ``beam" of the top K most likely sequences at each step. It not only to provides results with low error rates, but also allows for integration of external information, e.g., language models, contextual biasing, etc., that greedy search cannot account for.


\begin{figure}[h]
\includegraphics[scale=0.7]{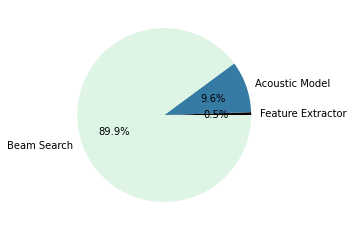}
\caption{Timing breakdown of NeMo's Conformer CTC Large inference pipeline on Lbirispeech test-clean, using Librispeech's 3-gram.pruned.3e-7.arpa.gz language model. Beam search ran on one CPU core, while feature extractor and acoustic model ran on one A100-80GB PCIe.}
\label{fig:timebreakdown}
\end{figure}

While major components of ASR pipelines are already GPU accelerated, e.g., feature extraction and acoustic model, current beam search implementations for CTC models are all CPU-based, consuming the majority of inference runtime, c.f., figure \ref{fig:timebreakdown}. Even though both feature extraction and acoustic model are already offloaded to the GPU, beam search on the CPU consumes nearly 90\% of the runtime. While there continue to be software-level innovations to accelerate (neural network) acoustic model inference on GPU \cite{dao2022flashattention}, Amdahl's Law limits end-to-end speed-up, unless beam search is also accelerated. The maximum speed-up that can be achieved by further accelerating the GPU part of the workload is a mere 11\%. 

Beam search can easily be parallelized over the elements in a batch and run on multiple CPU cores, alleviating the decoder bottleneck. However, performance does not scale linearly with the number of cores and the ratio of CPU cores per GPU in a system puts an upper bound on this approach. To date, there are no cloud systems with a ratio of CPU cores to GPUs exceeding 16. E.g., Nvidia's DGX A100 pairs 128 CPU cores with 8 A100 GPUs \cite{DGXA100} and AWS's EC2's p4d.24xlarge offers only 96 CPU cores with 8 A100 GPUs \cite{EC2P4}. The results in figure \ref{fig:timebreakdown} were obtained using the Flashlight Decoder on a single core. Even when enabling 16 CPU cores for decoding it remains the bottleneck consuming 43\% of end-to-end runtime. Our proposed GPU decoder (henceforth, ``CUDA WFST Decoder") alleviates the bottleneck, accelerating decoding (over the 16 CPU core baseline) by a factor of 15.6 and yielding a 4.6x end-to-send pipeline speed-up, c.f., section \ref{sec:offline}. 



Given that feature extraction and acoustic model are already GPU based, the end-to-end performance gains from using a GPU decoder directly translate into cost and power savings, the latter becoming a major concern for data centers. In other words, accelerating beam search is imperative for cost- and power-efficient ASR inference. 

The rest of the paper is organized as follows. We discuss related work and then describe algorithm and implementation details. We evaluate offline transcription performance and word error rate; real-time streaming transcription throughput and latency; and word boosting's effect on word error rate. We summarize our contribution and conclude with describing future work.




\section{Related work}

There are a few GPU-accelerated beam search decoders for ASR. Ours is an extension of the work described in \cite{hbraun}, which supports only hybrid DNN-HMM models in Kaldi. \cite{argueta-chiang-2017-decoding} lacks online decoding or ability to condition on acoustic model log likelihoods. \cite{DBLP:journals/corr/abs-1804-03243} has a critical error: it tries to synchronize CUDA thread blocks, which is disallowed by the CUDA programming model; this causes sporadic crashes. It also does not support streaming inference. K2 \cite{k2} also has a GPU-accelerated WFST-based CTC decoder, but it does not support online streaming inference or the other extensions we describe. All support n-gram language models.

The most popular CPU-based decoder today is Flashlight's \cite{kahn2022flashlight}. It is a non-WFST dynamic decoder that uses KenLM and prefix tries. It does support streaming inference. It is faster than other CPU-based decoders, but unfortunately not fast enough to keep the GPU fully fed doing acoustic model inference. It does not support word boosting. Its python bindings also do not support multi-threading because they do not correctly drop the global interpreter lock while running the C++-based decoder code. We forked the code base and added proper multi-threading support for the sake of fair comparison. It is the baseline implementation we compare to in this paper.




\section{Algorithm description}

\begin{figure}[t]
\centering
\includegraphics{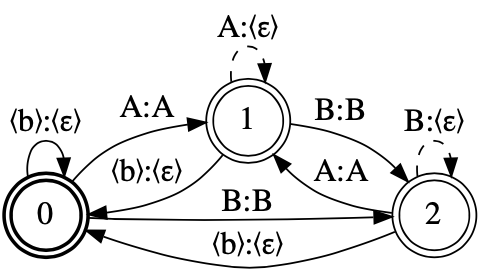}
\caption{``Normal" $T$ transducer that encodes the CTC transduction rules (that blank should not be emitted as an output, and that repeating the same token on the output requires at least one blank input in between). Adapted from \cite{Laptev_2022}.}
\label{fig:Tvanilla}
\end{figure}

\begin{figure}[t]
\centering
\includegraphics{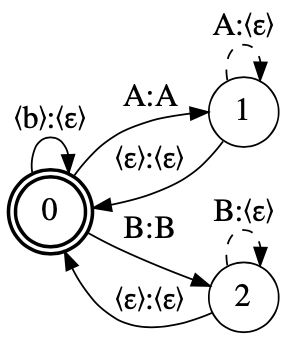}
\caption{``Compact" $T$ transducer. It does not encode the CTC rule that two consecutive symbols that are the same should be transduced to a single symbol. Adapted from \cite{Laptev_2022}.}
\label{fig:Tcompact}
\end{figure}

The exact details of the WFST decoding algorithm are described in detail in \cite{hbraun}. At a high level, an iteration is run for the log likelihoods output by the acoustic model at each timestep. There are three stages that run for each iteration: (1) ExpandArcs to expand the current set of best tokens to a new set of tokens, (2) Prune to remove tokens outside the ``beam", and (3) D2H Copy to copy these tokens and backpointers from the device to the host in order to construct a lattice or least cost path later.

This generic WFST decoder can be used for CTC models simply by creating a $T$ WFST that models the rules of CTC sequence transduction properly. $T$ is then composed with $L$ and $G$ WFSTs \cite{hbka}. For more details, see \cite{Laptev_2022}. Like \cite{Laptev_2022}, we investigate topologies that do not perfectly model the CTC transduction rules. In particular, we investigate the ``compact" $T$ WFST topology, finding that it greatly improves word error rate compared to the correct ``normal" $T$ WFST topology. See figures \ref{fig:Tvanilla} and \ref{fig:Tcompact}. For a discussion of the creation of the $TLG$ WFST used for the actual speech decoding, see \cite{hbka}. Our software library includes scripts to create these WFSTs.

We also make several improvements to the code in \cite{hbraun} to improve throughput without affecting the correctness of the algorithm:

\begin{enumerate}
    \item Since the creation of \cite{hbraun}, CUDA kernel execution has gotten faster by a factor of two just from hardware architectural improvements. This has caused kernel launch overhead to take up to 20\% of each iteration of beam search because it is often the case that the previous kernel has finished long before the CPU has launched the next kernel. We eliminate this CUDA kernel launch overhead by using cuda graphs \cite{cudagraphs}.
    \item The original implementation waited to do the next iteration of beam search until the previous iteration's tokens had been copied to CPU. Since device-to-host latencies have not improved commensurately with GPUs' architectural improvements, this takes up about 25\% of each iteration. Fortunately, memory capacity on GPUs have increased by a factor of two or more since then, so we can use double-buffering to prevent these stalls without worry about memory usage.
\end{enumerate}

We next discuss the particulars around implementing streaming inference and word boosting on top of this decoder.

\subsection{Streaming inference}

Effective streaming inference with GPUs requires dynamic batching of incoming streams of audio. In streaming automatic speech recognition, a new chunk from a given audio stream comes in at a nearly constant rate, typically at some multiple of the frame shift of the audio featurizer (frequently, but not always, 10 ms). Because the latency of doing inference on a given chunk is effectively always less than the this request rate, we must save the state of the acoustic model (e.g., the hidden state vectors of an LSTM), the state of the decoder (i.e., the current active tokens), and the state of the feature extractor (e.g., the random seed used for dithering and -rarely in modern architectures- the current ivector value) between requests from the same audio stream. While that audio stream is waiting for its next chunk, requests from other ready audio streams can be serviced. Fortunately, saving these states is effectively done by doing a scatter operation to a memory buffer in GPU memory. Meanwhile, loading can be done by a gather operation.

To help interpret our results, we make a basic observation from queueing theory \cite{Cahn1998}. We can model our inference server as an M/D/1 queue. That is, we have only one server, which takes a constant amount of time ($\mu = 1 / D$, the ``serving rate") to process each request; meanwhile, requests come in following a Poisson distribution at a rate of $\lambda$. Note that $D$ here refers to the latency of the server to respond to a request, often referred to as the ``compute latency". This is not a perfect model because serving rates have variance and this model assumes an infinite length queue (but computer memory is finite). A Poisson distribution is a reasonable model for incoming requests because each request is actually a batch of audio chunks from multiple audio streams. While the audio from a single stream comes in at an approximately constant rate, there is a variable wait time because earlier arriving audio streams must wait until a full batch worth of requests arrives.

In this model, the average total latency to service a request can be modeled as:

\[ \omega = \frac{1}{\mu} + \frac{\lambda}{2 \mu (\mu - \lambda)} \]

The first term corresponds to the time to process a request (compute latency), while the second term corresponds to the time a request spends waiting (queue latency). The main take away is that request wait time decreases quadratically with the serving rate $\mu$, because of the second term. Since serving rate is just the inverse of the compute latency $D$, we observe that the best way to decrease total latency for a request to get a response in streaming inference regimes is to decrease the latency of the inference engine, not necessarily to increase its throughput by scaling to larger batch sizes.

\subsection{Word boosting}

\begin{figure}[h]
\centering
\includegraphics[width=1.5cm]{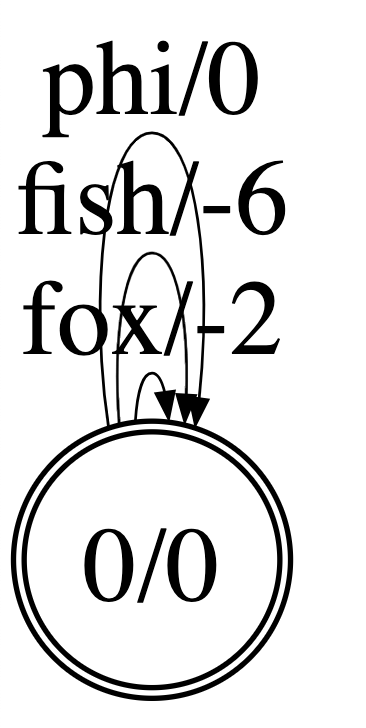}
\caption{Example word boosting WFSA. ``Fish" gets a boost of 6 and ``fox" gets a boost of 2. ``Phi" stands for a failure arc; if no other label is accepted, then that arc is taken.}
\label{fig:wordboostingfsa}
\end{figure}

We implement word boosting as follows. The decoder is given the WFST $TLG$ shared among all utterances. Then, when a new audio stream is created, the creator of it can also pass a word boosting graph $B$ via a table mapping in-vocabulary word IDs to scores. $B$ is created from the table as a single state Weighted Finite State Acceptor (WFSA); see figure \ref{fig:wordboostingfsa}. The decoder will then decode that specific utterance as if its decoding graph were $TLG \circ B$. Of course, $TLG$ is not explicitly composed on the right with $B$, since that would require making a copy of $TLG$ for every utterance requesting word boosting (which is typically several gigabytes in size). Instead, we take advantage of the fact that $B$ is a single state to implement it as a hash table on GPU (in particular, we use cuco::static\_map \cite{cuco}) mapping word IDS to scores. Every time an arc with a non-epsilon olabel is encountered in $TLG$ during the ExpandArcs step of the algorithm, we always know we will be in $B$'s one and only state, so we can simply check whether that olabel is in the cuco::static\_map object for that utterance. The pseudo-code is given in figure \ref{alg:wordboostexpandarcs}.

\begin{figure}[h]
\begin{algorithm}[H]
\small
\begin{algorithmic}[1]
\For{token $\in$ main\_queue}
\For{arc $\in$ token.state.out\_arcs}
   \State  new\_cost $\gets$ token.accumulated\_cost + acoustic\_cost[arc\_id.ilabel] + arc.graph\_cost
   \If{\colorbox{green!15}{arc\_id.olabel $!= 0$}}
   \State    \colorbox{green!15}{new\_cost $\gets$ new\_cost + b\_fst[arc\_id.olabel].graph\_cost}
   \EndIf
\EndFor
\EndFor
\end{algorithmic}
\caption*{\small \textbf{Pseudo-code of Expand Arcs operation to support word boosting. }}
\end{algorithm}
\caption{The changes to the original algorithm to support word boosting are highlighted in green. Note that in practice both for loops are run in parallel on the GPU.}
\label{alg:wordboostexpandarcs}
\end{figure}

Note that, since the decoder is seeking a minimum cost path, ``boosted" words should be given a negative score.

There are a few limitations to this approach. 

First, the ideal value for word boosting is dependent upon the beam width. Suppose that the numerical beam width is $W$, and that the word ``cat" has a boosting score of $W$ as well. Assuming that the log likelihood scores from the acoustic model and the $TLG$ arc scores are approximately equal at each time step, sequences containing words other than ``cat" will fall out of the beam, resulting in the final transcript almost always being a sequence of the word ``cat" repeated multiple times.

Second, $TLG \circ B$ will not be ``stochastic". That is, the arcs entering any given state will not necessarily sum to one in the log semiring. This does not appear to be an issue in practice. Note that, if following the WFST creation recipe used by Kaldi \cite{kaldi_graph},  $TLG$ will not be stochastic to begin with.

Third, this approach can work poorly if the olabels are not pushed as far left as possible in $TLG$. One can do this either by construction in the lexicon following the recipe in \cite{hbka} or by doing label pushing on the olabels of the lexicon in \cite{dmake}. This allows the boost scores to participate in beam search earlier rather than later. The scripts for graph creation in our open source software handle this quirk.

Fourth, out-of-vocabulary words will not be boosted because they naturally won't compose with the olabels of $TLG$.

\section{Results}

\begin{table*}[t]
\begin{tabular}{llrrrr}
\toprule
                   model & dataset &  flashlight wer\% &  cuda compact wer\% &   cuda normal wer\% \\
\midrule
   stt\_en\_conformer\_ctc\_small &   test-clean &        3.28 &            \textbf{3.17} &             3.56 \\
   stt\_en\_conformer\_ctc\_small &   test-other &        6.99 &            \textbf{6.83} &             7.12 \\
  stt\_en\_conformer\_ctc\_medium &   test-clean &        \textbf{2.59} &            2.65 &             3.20 \\
  stt\_en\_conformer\_ctc\_medium &   test-other &        5.43 &            \textbf{5.32} &             5.73 \\
  stt\_en\_conformer\_ctc\_large &   test-clean &        2.31 &            \textbf{2.21} &             2.74 \\
   stt\_en\_conformer\_ctc\_large &   test-other &        4.26 &            \textbf{4.10} &             4.53 \\
\bottomrule
\end{tabular}
\caption{Offline Word Error Rate Table. Best results per row are bolded.}
\label{tbl:wertable}
\end{table*}

\begin{table*}[t]
\begin{tabular}{lllrrr}
\toprule
                  model & dataset &  flashlight &  cuda compact &   cuda normal  & compact to flashlight speed-up \\
\midrule
  stt\_en\_conformer\_ctc\_small &   test-clean &         646 &          4320 &           \textbf{4464} & 6.7 \\
  stt\_en\_conformer\_ctc\_small &   test-other &         614 &          4110 &           \textbf{4528} & 6.7 \\
  stt\_en\_conformer\_ctc\_medium &   test-clean &         549 &          \textbf{3727} &           3721 & 6.8  \\
  stt\_en\_conformer\_ctc\_medium &   test-other &         536 &          3910 &           \textbf{4082} & 7.3  \\
   stt\_en\_conformer\_ctc\_large &   test-clean &         488 &          2261 &           \textbf{2322} & 4.6  \\
   stt\_en\_conformer\_ctc\_large &   test-other &         465 &          \textbf{2205} &           2189 & 4.7 \\
\bottomrule
\end{tabular}
\caption{Offline RTFx end-to-end pipeline throughput for flashlight, cuda compact, and cuda normal decoders}
\label{tbl:throughputtable}
\end{table*}


\begin{table*}[t]
\begin{tabular}{llrrr}
\toprule
                  model & dataset & Flashlight (single core) &  Flashlight (16 cores) &  cuda compact \\
\midrule
  stt\_en\_conformer\_ctc\_small &   test-clean & 1 &  4.1 &  47.5 \\
  stt\_en\_conformer\_ctc\_small &   test-other & 1 &  4.1 & 55.4 \\
  stt\_en\_conformer\_ctc\_medium &   test-clean & 1 & 4.6 & 67.0 \\
  stt\_en\_conformer\_ctc\_medium &   test-other & 1 & 4.4 &  66.6 \\
   stt\_en\_conformer\_ctc\_large &   test-clean & 1 & 3.4 &  53.2 \\
   stt\_en\_conformer\_ctc\_large &   test-other & 1 & 3.3 &  56.7 \\
\bottomrule
\end{tabular}
\caption{Speed-up factor of 16 core Flashlight Decoder and CUDA WFST Decoder with compact topology over single core Flashlight decoder on model-dataset pairs studied in table \ref{tbl:throughputtable}.}
\label{tbl:proportions}
\end{table*}

\subsection{Offline inference}
\label{sec:offline}


For offline evaluation, we compare word error rate and throughput for several setups. We evaluated three variants of NeMo's Conformer CTC architecture -  Conformer CTC Small (13 million parameters), Conformer CTC Medium (30 million parameters), and Conformer CTC Large (120 million parameters) - on Librispeech test-clean and test-other. We evaluated in three scenarios: (1) Using the Flashlight Decoder using 16 CPU cores, (2) using the CUDA WFST Decoder with the ``compact" CTC topology to construct $TLG$, and (3) using the CUDA WFST Decoder with the ``normal" CTC topology to construct $TLG$. We used the Librispeech 3-gram.pruned.3e-7.arpa.gz language model in all scenarios. A single A100-80GB PCIe card was used to run the feature extraction, acoustic model, and - except in the Flashlight Decoder case - beam search. We ran at batch size 200. Further hyperparameters settings are discussed below.

One area where the CUDA WFST Decoder and the Flashlight Decoder vary in terms of hyperparameters is their concept of maintaining a list of the current most likely hypotheses. Both have a concept of ``beam width"; hypotheses with a score gap from the best hypothesis larger than this floating point threshold are removed. We set this value to 17.0 for both decoders. However, they have different concepts for specifying the top number of hypotheses allowed at any given time. The CUDA WFST Decoder allows only a maximum number of hypotheses (called ``max active") that are defined in terms of the states of the WFST. The Flashlight Decoder defines its maximum number of hypotheses (called ``beam size") in terms of non-emitting CTC symbols. These are not the same because the WFST can have states entered via epsilon arcs, and because the Flashlight Decoder does not consider two sequences varying only in terms of blanks to be different. We set the max active for the CUDA WFST Decoder to 10,000 while the beam size for the Flashlight Decoder is 2,500. This choice reflects the differences discussed above.

Surprisingly, the compact WFST consistently achieves a better word error rate that the normal WFST. The hypothesized reason for this is that fewer tokens are used to occupy different states that end up leading to the same output string, since the ``deduplication" paths mapping non-blank and non-epsilon ilabels to blank olabels no longer need to be followed. The compact WFST topology also does better than the Flashlight Decoder in terms of word error rate, though the difference is smaller. This is likely because language model weights are pushed left in the WFST graph and thus able to inform search before a word is fully seen, an optimization not possible in dynamic decoders. Reference table \ref{tbl:wertable}.

We document the throughputs achieved when running the end-to-end pipeline, including feature extraction, acoustic model inference, and beam search in table \ref{tbl:throughputtable}. According to table \ref{tbl:throughputtable}, the compact and normal topologies dominate the Flashlight Decoder in terms of throughput, measured in terms of RTFx (hours of audio transcribed per hour of wallclock time). Given the lower word error rate of the compact CTC topology, we compute the throughput ratio of the CUDA WFST Decoder with the compact topology to the Flashlight Decoder in the final column. The throughput ratio is between 4.63 and 7.29, depending on model size and dataset. Model size affects throughput because the acoustic model inference time takes longer as models are larger. Slightly more time is spent in beam search when running on test-other than on test-clean because the acoustic model outputs less confident predictions. At a high level, this means that fewer tokens can be pruned, which means more more data to copy in the D2H Copy step and more work in the next iteration's ExpandArcs.

In order to elucidate the relationship between the end-to-end pipeline throughput increase in table \ref{tbl:throughputtable} and the throughput increase of the beam search decoder itself, we show the relative speed-up of the Flashlight Decoder using 16 CPU cores and the CUDA WFST Decoder vs. the Flashlight Decoder with a single CPU core on only the beam search part of the pipeline in \ref{tbl:proportions}. The Flashlight Decoder has poor scaling, achieving ~4x speed-up with 16 cores instead of closer to 16x. Even if Flashlight achieved perfect scaling, the CUDA WFST Decoder is still several times faster.

Thus, we can conclude that the CUDA WFST Decoder, used with the compact CTC topology, expands the Pareto frontier in terms of throughput-accuracy tradeoffs, compared to the Flashlight Decoder.

\subsection{Streaming inference}

We evaluated latency performance in the context of a 4 layer latency-controlled BLSTM \cite{7953176}. The BLSTM had a right hand context of 20 10ms frames, and had four BLSTM layers, with a hidden size of 600 and an internal projection size of 300. We used the Hub5'00 corpus of phone conversations to simulate a real-time inference scenario for real-time video conferencing subtitling. Each request to the inference server corresponded to 600ms of audio data, or 60 frames, each 400ms apart.

Our benchmark was implemented via a Triton \cite{triton} Python backend, taking advantage of Triton's ability to queue and dynamically batch data from multiple audio streams. We sent data at a constant rate of 2.5 requests per second (corresponding to sending 2.5 400ms windows of audio every second), with 256 incoming audio streams per second. We benchmarked two configurations: (1) Doing feature extraction and acoustic model inference on the GPU, but beam search decoding on the CPU with the Flashlight Decoder; (2) Doing all parts of inference on the GPU, using the CUDA WFST Decoder.

Results are summarized in table \ref{tbl:online}. The ``Total Latency" is simply the sum of the compute latency and the queue latency. For a full picture of latency, one would need to add in the network latency and right hand context latency (200ms in this case). Nevertheless, the table shows the disastrous effect of queue latency's quadratic dependency on compute latency. The version using the Flashlight Decoder on CPU has a 430ms average response time, while the CUDA WFST Decoder version achieves 54.4ms average response time, despite having less than half the latency of the Flashlight Decoder. To make this concrete, the average syllable duration is 100ms, so the Flashlight Decoder is over four syllables late, while the CUDA WFST Decoder is only half a syllable late. Furthermore, the Flashlight Decoder cannot keep up with incoming data: it achieves RTFx of 235.4, while the maximum is 256 RTFx (because 256 audio streams are coming in in real time). This means that the queues are constantly growing in memory usage the longer the server runs. The only way to remedy this is to make the server serve fewer audio streams (e.g., 128). Word error rate between both decoders was equal. Therefore, we can conclude that for online inference as well as offline inference, doing beam search on the GPU is a requirement for full utilization of hardware.

\begin{table*}[t]
\begin{tabular}{lrrrrrr}
\toprule
                  Decoder & Avg Compute Latency & Avg Queue Latency &  Avg Total Latency & P99 Total Latency & RTFx \\
\midrule
  CUDA Compact Topology & 30.1 & 24.4 & 54.5 & 78 & 256.0 \\
  Flashlight & 53.7 & 376.2 & 430.0 & 541.5 & 235.4 \\
\bottomrule
\end{tabular}
\caption{Online Latencies and throughputs. All units are in milliseconds, except for RTFx, which has no unit.}
\label{tbl:online}
\end{table*}

\subsection{Word boosting}

\begin{figure}[t]
\includegraphics[width=9cm]{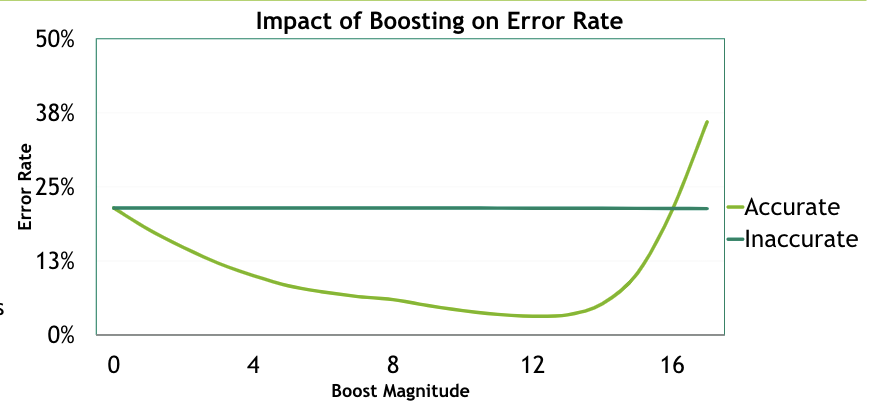}
\caption{Effect of Boost Magnitude on Word Error Rate}
\label{fig:wordboosting}
\end{figure}

Figure \ref{fig:wordboosting} shows the word error rate of a Conformer CTC Large model with an internal language model with a beam width of 32 evaluated on an internal 5 hour long low signal-to-noise ratio test dataset in two scenarios:

\begin{enumerate}
    \item
    The ``Accurate" scenario where every unique word in the test dataset is boosted by (the negation of) ``Boost Magnitude"
    \item
    The ``Inaccurate" scenario where only words that are not in the test set (but are in the vocabulary of the $TLG$ model) are boosted by (the negation of) ``Boost Magnitude". The number of out-of-vocabulary words boosted is equal to the number of unique words in the test set.
\end{enumerate}

The fact that the ``Inaccurate" curve is a straight line shows that, for reasonable boost scores, there is no disadvantage to word boosting in terms of accuracy. Note that, in the ``Accurate" scenario, word boosting helps until the boost magnitude reaches about half the beam width, because of the phenomenon discussed above. Therefore, as a rule of thumb, we recommend using a boost value of at most half the beam width.
Decrease in throughput and increase in latency are miniscule; therefore, they are not depicted.

\section{Conclusions and Future Work}

To the best of our knowledge, this is the first work that investigates how to fully utilize hardware acceleration for end-to-end automatic speech recognition in both offline and online scenarios. We achieve between 4.6 and 7.3 times greater throughput on the same server hardware compared to the baseline. In cloud environments where developers can dynamically provision instances with fewer CPUs, this will reduce cost further than the 4.6 to 7.3 figure.

As far as future work is concerned we plan to extend word boosting support to enable out-of-vocabulary words or boosting phrases using techniques described in \cite{44202}.


Last, we'd like to note, that several of the techniques described here (dynamic batching for streaming inference, word boosting, etc.) can be applied on top of K2's CUDA-based RNN-T decoder \cite{k2decoder}. This provides a path towards high throughput, low latency inference for RNN-T models which are gaining popularity.

\bibliographystyle{IEEEbib}
\bibliography{refs}

\end{document}